# Artificial intelligence (AI) techniques: a game-changer in Digital marketing for shop

Suzan Abbas Abdullah

Technical College Baghdad, Middle Technical University, Iraq

Suzan.abbas@mtu.edu.iq

techcolbag@mtu.edu.iq

https://orcid.org/0000-0002-9661-4802

**ABSTRACT**

The quick growth of shops using artificial intelligence (AI) techniques has changed digital marketing activities and changed how businesses interact and reach their consumers. (AI) techniques are reshaping digital interactions between shops and consumers interact digitally by providing a more efficient and customized experience, fostering deeper engagement and more informed decision-making. This study investigates how (AI) techniques affect consumer interaction and decision-making over purchases with shops that use digital marketing. The partial least squares method was used to evaluate data from a survey with 300 respondents. When consumer engagement mediates this relationship, artificial intelligence (AI) techniques have a more favorable impact on purchasing decision-making. Consequently, decision-making is positively impacted through consumer engagement. The findings emphasize that for a bigger impact of the (AI) techniques on decision-making, the consumer must initially interact with the (AI) techniques. This research unveils a contemporary pathway in the field of AI-supported shop engagements and illustrates the distinct impact of (AI) techniques on consumer satisfaction, trust, and loyalty, revolutionizing traditional models of customer-purchase decision-making and shop engagement processes. This study provides previously unheard-of insight, into the revolutionary potential of (AI) techniques in influencing customer behavior and shop relationships.

1. **INTRODUCTION**

Artificial intelligence is widely viewed as a system that exhibits intelligent behavior by evaluating the environment and executing appropriate actions Sheikh, H, (2023).The significance of artificial intelligence (AI), a rapidly developing technique, on marketing is undeniable. Technique that makes machines or computers as intelligent as humans Hermann, E. et al., (2024). The ability to execute is linked to actions that resemble those carried out by the human brain.  It has been used in conjunction with digital marketing to help firms more easily reach Consumers at the appropriate moment. As stated by Joshi, S et al. (2025)

 In response to this shift, new techniques have unlocked a vast realm of opportunities, enabling shops to enhance operational efficiency and deepen consumer relationships Ravić et al., (2022).

Machine learning (ML) and artificial intelligence (AI) techniques stand out among the new techniques because they give marketers the ability to monitor, understand, and forecast consumer behavior on-demand. By grouping their consumers according to their demographics, habits, and interests, shops can also employ AI techniques to automatically create consumer profiles Kempton (2023).

Furthermore, sophisticated skills that can replicate and in certain situations, exceed human performance are introduced by AI techniques. These technologies have the potential to drastically alter digital marketing tactics and boost sales in stores of all sizes. The creation of effective marketing plans is crucial to this change and is still necessary for success Almaiah et al., (2022).

Shops must keep up with digital marketing innovations in today's fiercely competitive industry in order to effectively satisfy changing consumer demands Arantes, (2023).

Shops are fighting to get in and keep customers while outsmarting one other because consumers today have a lot of options. Shops use a range of strategies to compete with their rivals, including cost leadership, uniqueness, or both. This leads to market penetration, segmentation, and a host of other results. How stores can benefit is the question. Understanding competitors' capabilities and staying ahead of them is the key to success; this is known as marketing intelligence Ghazi et al., (2023).

AI techniques are a potent tool in this context, allowing stores to assess consumer attitudes and emotions on a large scale. Shops can discover new consumer trends, make well-informed strategic decisions, and improve consumer service techniques by using AI-driven sentiment analysis. This will ultimately increase consumer pleasure, loyalty, and trust. Bozkurt S et al., (2022).

As a result, it's critical to highlight how AI techniques may improve marketing decisions and increase productivity, encourage scientific research, establish a department specifically focused on using AI to activate digital marketing in shops, and Shops can increase marketing effectiveness and enhance the consumer experience in line with global trends by utilizing AI techniques Jayasingh, S et al.,( 2025).

Consumer engagement with AI techniques has been recognized as a crucial element in predicating and understanding consumer behavior outcomes, including Consumer purchasing decision-making Hollebeek, L.(2011). Engaged customers on shops pages usually exhibit a positive attitude towards the shops and also show higher Consumer purchasing decision-making to buy Wang, R.J.-H, (2020).

Consequently, we will also utilize purchase decision-making as our outcome variable to examine how consumer engagement Customer' decision-making to make purchase .Accordingly, this study aims to the growing body of knowledge on AI techniques role in digital marketing by presenting a structured framework that integrates the AI techniques, marketers, and consumer engagement and purchasing decisions, as well as the consumer's viewpoints. This lead to two central research questions:

- What is the impact of AI techniques on purchasing decision-making with shops utilizing AI in Iraq?

- How will AI techniques-based Digital marketing strategies, influence customer behavior?

To address these questions, 300 responses from Iraqi consumers were collected and examines mediating role of consumer engagement in the relationship between to AI techniques and consumer decision-making. By doing so, it advances understanding of how AI -powered tools shape consumer behavior and supports the strategic application of AI within digital marketing contexts. The study structure includes the following: Section 1 introduction. Section 2 reviews the relevant literature. Section 3 Develop and create hypotheses. Section 4 Research Method. Section 5 Results. Section 6 discusses. Section 7Conclusion.

## 2. Literature review

### 2.1. Digital marketing

Digital marketing as a dynamic, technology-enabled process by which businesses collaborate with customers and partners to jointly create, deliver, and sustain value for all stakeholders. As defined by Kannan and Li (2017). Its evolution was notably accelerated by the birth of the World Wide Web in 1990. Which laid the foundation for a transformative shift in marketing practices.

According to Akgun et al., )2021),and in line with the Digital Marketing Institute (DMI) digital marketing focuses on using digital technologies to establish meaningful and measurable and integrated relationship with customers. The ultimate aim is to acquire, retain, and deepen Consumer engagement through data-driven strategies.

As Şentürk et.al. (2023) observe in earlier years, firms had a penchant for employing digital marketing methods was a matter of preference. Businesses now need to be on digital media and employ digital marketing techniques, especially after the events such as the COVID-19 pandemic, which has impacted and limited the entire world, wars, and applications that have caused the loss of in-person engagement with customers.

Joshi (2021) argues that the last ten years have seen a rapid adoption of modern technology, making customers impatient for anything that can be had quickly, easily, and affordably. Because of this, practically all businesses have had to go through the digitization phase in order to become more automated, productive, and evolving. To provide high-quality services, digital marketing is reshaping consumer and industry behaviors.

In the view of Tam, F. Y., and Jane Lung (2025), as well as Sutherland, K (2025), digital marketing is actively reshaping not only businesses practices but also consumer behavior, reflecting a broader shift in how value is created and consumed in the digital economy.

Moreover, Ma X, Gu X. (2024) point out check Small business owners that can no longer withstand economic hardships have given up on physical shops and traditional marketing strategies, and many have only lately started to compete in the online market.

As Kotler & Armstrong (2020) explain the digital marketing entails using technologies like websites, social media, mobile ads and applications, online video, email, blogs, and other digital platforms, additionally, it entails using computers, smartphones, tablets, internet-ready TVs, and other digital devices to interact with customers at any time and from any location.

Ma X, Gu X. (2024) explained there are many different digital marketing strategies (content marketing, social media marketing, search engine marketing) and marketing tools (data analytics, search engine optimization, and marketing automation) that influences a mobile app or web site. Da Silva, D et al., (2023).emphasize marketing any strategy that can significantly influence people at a certain time, location, and through a specific channel.

To sum up, digital marketing has changed how companies interact with increasingly tech-savvy customers, moving from being a supplemental tool to a strategic need. It allows businesses to react quickly to changes in the market and in the behavior of their customers through personalized and real-time engagement. This fundamental change lays the groundwork for using cutting-edge strategies, especially AI strategies, to improve marketing accuracy and consumer focus**.**

## 2.2. Intelligent machines

The phrase "artificial intelligence" was coined by John McCarthy in 1956. He defined it as "the science and engineering of making intelligent machines, especially intelligent computer programs" and added that it was comparable to the similar goal of using computers to study human intellect, though AI need not be restricted to physiologically observable methods.

As noted by Bagnoli C, et al., (2022) the development of artificial intelligence (AI) has drastically altered how industries function, substituting automation and digital solutions for human labor and conventional techniques. AI makes it possible for robots to make wise decisions, which boosts productivity and creativity. The development of automation and robotics in the second half of the 20th century is where artificial intelligence got its start. With the introduction of machine learning and deep learning technologies in the 2010s, however, a real revolution took place. These technologies have made it possible for industries to successfully address difficult issues.

According to Gao Y. et al. (2025), the usage of artificial intelligence is characterized by people using AI systems to evaluate and learn from outside data and accomplish predetermined objectives and tasks through ongoing improvements.

Although many writers have discovered different approaches to AI techniques, we have chosen to concentrate on a select handful for the purposes of this work. A synopsis of the primary techniques is provided below:

Natural Language Processing (NLP): Aims to make it possible for computers to understand, generate, and interpret human language in order to gain practical understanding in order to address problems. NLP seeks to accomplish two objectives: (1) extract information from written language; and (2) facilitate human communication (I. Taboada, et al.(2023).

Machine Learning (ML): goal of create algorithms and models that let computers carry out tasks without direct programming guidance. Machine learning algorithms are made to draw conclusions and learn from data patterns, which enables them to evolve and perform better over time. Supervised learning, unsupervised learning, and reinforcement learning are some of the methods used in this subject. The goal of machine learning is to enable machines to make decisions on their own L. Sharma and P. K. Garg, (2021).

Neural Networks (NNs): these are computational models inspiration for neural networks (NNs). They consist of interconnected nodes organized in layers. NNs use layers of connected neurons to interpret input data. Each neuron executes an activation function after performing a weighted sum of its inputs. Sule, S., (2025)

Fuzzy logic (FL): is a mathematical method that deals with reasoning that is approximate rather than exact, enabling the representation of ambiguity and uncertainty in decision-making. By concentrating thought and decision-making on all possible answers between 0 and 1, FL extends Boolean logic (1 or 0). (I. Taboada and others, 2023)

Expert Systems (ES): are artificial intelligence (AI) programs that simulate a human expert's decision-making process in a certain field. These systems employ inference engines to draw conclusions and offer suggestions or answers after encoding domain-specific knowledge as facts or rules. Expert systems provide consistent and dependable decision support by capturing and preserving knowledge and experience in particular fields L. Sharma and P. K. Garg, (2021).

Chatbots: are artificial intelligence (AI) systems created to mimic a natural language conversation with a user. It can communicate over the phone, through mobile apps, websites, and messaging services. Furthermore, it is possible to link chatbots with other apps. Cubric, M. (2020)

Collectively, these AI techniques represent the shift from theoretical research to real-world implementation, providing a strong basis for the digital era. Together, the examined literature shows how AI approaches support better decision-making, increased productivity, and the creation of intelligent systems that support societal and organizational objectives.

## 2.3. Artificial Intelligence in Marketing

Artificial Intelligence (AI) has emerged as a pivotal tool in transforming digital marketing. According to Shaik (2023), AI plays a crucial role in organizing and interpreting vast and diverse data collected from multiple digital points of sale. This enables marketers to make swift, data-driven decisions, deliver personalized consumer service, and streamline marketing operations, thus gaining a competitive edge. AI is increasingly seen as a revolutionary force in the realm of digital marketing.

Krishna et al. (2023) emphasized that AI techniques empowers businesses to collect and monitor real-time, accurate consumer data. This data can be leveraged to craft highly personalized and effective marketing strategies tailored to specific consumer behaviors and preferences.

Todorova and Antonova (2023) examined various AI techniques in marketing, including content creation, audience targeting, process optimization, sales enhancement, and chatbot implementation. They argue that AI represents a central driver of marketing process efficiency and highlight its potential in fostering AI-human collaboration as a key aspect of future marketing ecosystems.

The broader integration of AI across enterprises is widely recognized for its potential to significantly boost competitive advantage. Cooper (2021) and Agarwal et al. (2023) indicate that AI's value lies primarily in its support for lean and agile product development methodologies. These approaches aim to minimize waste and enhance efficiency, while AI complements them by shortening innovation cycles, detecting development issues in real time, and providing actionable improvement insights.

Babina et al. (2024) support this by noting that AI techniques facilitates real-time data analysis, helping to identify potential challenges and offering improvement strategies, thereby reducing resource waste and accelerating innovation.

Moreover, AI techniques are instrumental in analyzing user feedback regarding product functionalities shared online (Kushwaha et al., 2021). Such analysis enables businesses to better understand consumer needs and make timely enhancements. AI algorithms are also employed to study consumer preferences and recommend suitable products and services. In this context, algorithmic pricing becomes a powerful tool, offering two primary models: dynamic pricing, which adjusts prices based on demand and situational factors, and personalized pricing, which customizes prices for individual customers based on their specific preferences and behavior (Seele et al., 2021).

Raneri et al. (2023) further emphasize that AI techniques can rapidly identify functional requirements for new product versions, thereby expediting the "build–test–modify" cycle. This agility enables businesses to respond promptly to market dynamics and improve both adaptability and competitiveness.

Huang and Rust (2021) highlight AI techniques's growing significance across broader marketing domains. They note its application in various roles, such as consumer service robots, big data analytics for pricing and forecasting, recommendation engines for personalized marketing, natural language processing (NLP) for client interaction, and in-store analytics for monitoring consumer satisfaction.

Furthermore, Hu et al. (2021) underscore AI techniques's dual function: interacting with consumers through natural language while simultaneously gathering valuable data. This dynamic encourages customers to share personal insights in exchange for tailored solutions and recommendations (Whang & Im, 2021), reinforcing the cycle of data-driven personalization in modern marketing strategies.

**2.4 Research Contribution and Novelty**

Despite the significant advancements in the application of artificial intelligence (AI) techniques in digital marketing globally, prior studies have largely overlooked the specific impact of AI on consumer purchasing decisions within the context of digital shops in Iraq. This market presents unique cultural and economic characteristics that differentiate it from other regions. This study addresses this gap by employing Partial Least Squares Structural Equation Modeling (PLS-SEM) to examine the mediating role of consumer engagement with AI techniques an approach seldom applied in similar regional research. By integrating AI techniques, consumer engagement, and purchasing decision-making into a comprehensive framework, this research not only advances the theoretical understanding but also provides practical insights for digital retailers aiming to enhance marketing effectiveness in emerging markets. Consequently, it contributes to enriching the academic literature and supports the strategic implementation of AI in underexplored contexts.

**3- Develop and create hypotheses**
**First, the connection between AI techniques and consumer engagement in the shops that use digital marketing**

Artificial Intelligence (AI) techniques have revolutionized digital marketing strategies by facilitating personalized, real-time, and data-driven interactions with customers. Kushwaha et al., (2021). These techniques are believed to promote consumer engagement by increasing satisfaction, fostering trust, and encouraging loyalty toward businesses that implement them in their digital marketing strategies Bozkurt S et al., (2022). Thus, the following theory was developed:

• H1a: The application of AI techniques has a significant positive effect on consumer satisfaction in shops utilizing digital marketing.

• H1b: The application of AI techniques has a significant positive effect on consumer trust in shops utilizing digital marketing.

H1c: The application of AI techniques has a significant positive effect on consumer loyalty toward shops utilizing digital marketing.

**Second. The impact of AI techniques on consumer purchasing decision-making in shops utilizing digital marketing**

AI techniques driven marketing efforts offer tailored recommendations, predictive analytics, and efficient communication, all of which may play a critical role in shaping consumer decision-making processes. Understanding how AI affects consumers' purchase decisions is essential for evaluating its effectiveness in digital marketing environments. Sharma, A.P et al., (2024). Consequently, the ensuing hypothesis was proposed:

• H2: AI techniques exert a positive and significant influence on consumers' purchasing decision-making processes in shops utilizing digital marketing.

**Third. The influence of consumer engagement with shops utilizing digital marketing on Consumer purchasing decision-making**

Consumer engagement, reflected through satisfaction, trust, and loyalty, is a key driver of purchasing behavior. When customers interact with AI-powered systems that meet their expectations, their engagement is likely to influence their purchasing decisions positively, especially in the subject of digital marketing services Gomes, S et al., (2024). Consequently, the following hypothesis was created:

• H3a: Consumer satisfaction with AI techniques driven shops positively influences purchasing decision-making in the subject of digital marketing.

• H3b: Consumer trust in AI techniques driven shops positively influences purchasing decision-making in the subject of digital marketing.

• H3c: Consumer loyalty toward AI techniques driven shops positively influences purchasing decision-making in the context of digital marketing.

**Fourth. The mediating role of consumer engagement with shops utilizing digital marketing confirms the positive connection between AI techniques on consumer purchasing decision-making**

AI techniques directly impact purchasing decisions, this relationship may be better understood by examining the mediating role of consumer engagement. Satisfaction, trust, and loyalty may serve as important mechanisms through which AI-driven marketing efforts affect consumer choices Pelau C, et al., (2021). Consequently, the subsequent hypothesis was formulated:

• H4a: The positive effect of AI techniques on purchasing decision-making in shops utilizing digital marketing services is mediated by consumer satisfaction.

• H4b: The positive effect of AI techniques on purchasing decision-making in shops utilizing digital marketing services is mediated by consumer trust.

• H4c: The positive effect of AI techniques on purchasing decision-making in shops utilizing digital marketing services is mediated by consumer loyalty. Then, the following hypothesises was posited in (Fig. 1).

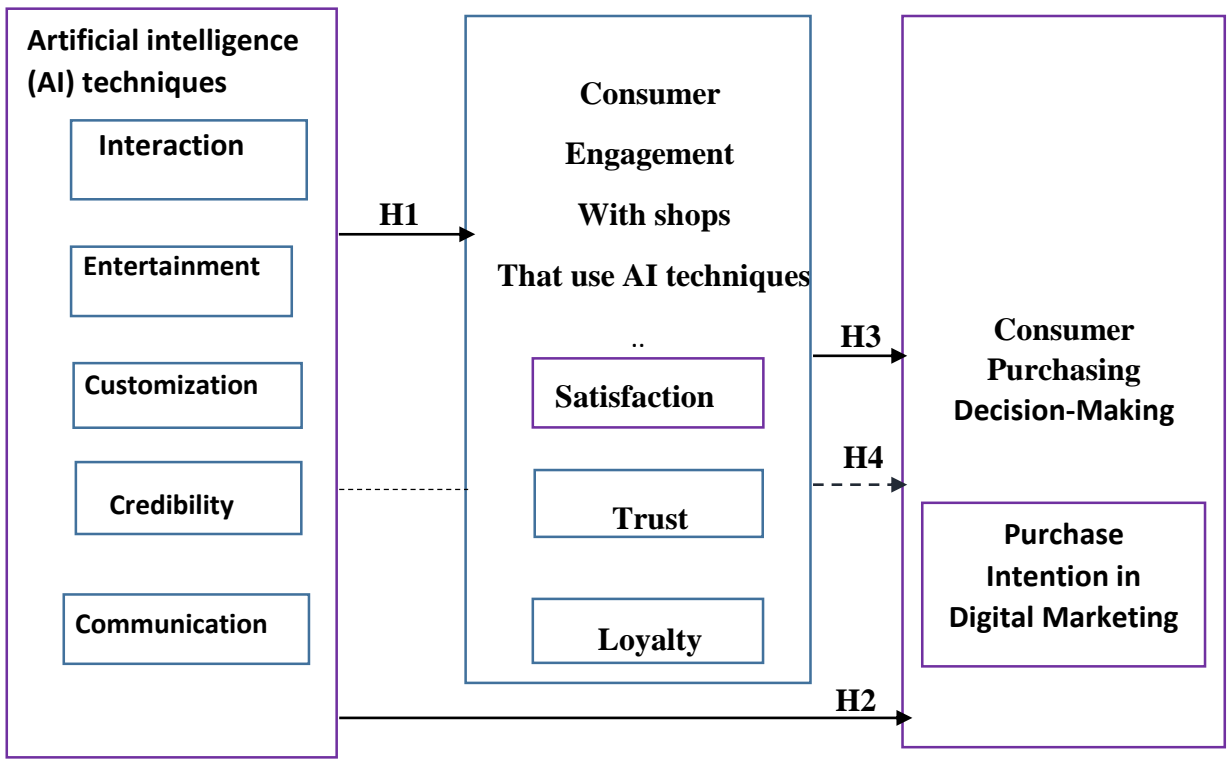

**NOT:** Direct factors ( ⟶ ) and mediating factors ( ----▶ )
**Fig. 1 Research model and hypotheses.**
**4- Research Method**
**4.1 Measures and the Sample**

The information used in this study was gathered during February and March 2025 utilizing a survey conducted online, the link to which was shared via the author's social networks and personal contacts. Due to practical constraints including accessibility and time, a convenience sampling method was employed to select participants for this study, while this sampling technique limits the generalizability of the findings to the broader population, the sample size and targeted criteria provide valuable initial insights into consumer behavior. The target population comprised Iraqi consumers, the criteria for participation required that respondents: (i) be a resident of Iraq and at least eighteen years old. And (ii) have bought at least one item. From a shop utilizing artificial intelligence (AI) techniques within the past year. A total of 300 valid responses were obtained, forming the final sample that fully satisfied these conditions.

Participation was unnamed, and informed consent was secured from all respondents prior to questionnaire completion. To ensure clarity from the items and determine the mean answer time, a preliminary experiment was conducted with a pilot group of 10 participants. The results of this pilot indicated an average completion time of approximately five minutes, with no reported comprehension difficulties.

The questionnaire comprised three main sections (detailed in Appendix A) adapted from Gomes and Lopes (2025). In the first section, respondents were instructed to:

"Think of a product purchased from an online store that uses AI techniques. Answer the following questions based on this experience."

This part measured five AI-related constructs: (i) Interaction (3 items), (ii) Entertainment (4 items), (iii) Customization (4 items), (iv) Communication (3 items), and (v) Credibility (3 items).

The second part assessed consumer interaction with online shops employing AI, consisting of three constructs: four items for satisfaction, five items for trust, and four items for loyalty.

The final part measured Consumer Purchasing Decision-Making concerning AI-driven shops using three items. All items across the questionnaire were evaluated on a 5-point Likert scale ranging from 1 (Strongly Disagree) to 5 (Strongly Agree), with the exception of purchasing decision-making items. These were assessed through three distinct 5-point Likert sub-scales: 1 (Improbable) to 5 (probable); 1 (Impossible) to 5 (Possible); 1(Dislike) to 5 (Like).

## 4.2 Research methodology

The methodology used in this study was quantitative. Initially, using SPSS software (v. 25), a statistical analysis was conducted on the items measuring the constructs. Following this, the items were then categorized into factors using an Exploratory Factor Analysis (EFA), and factor loadings and the reflective nature of the study methodology were evaluated using a Confirmatory Factor Analysis (CFA). Then, using Smart PLS (v. 4.0) software, the Partial Least Squares (PLS) technique was used to the study strategy. This approach integrates regression estimation with factor analysis based on the Ordinary Least Squares (OLS) method.

When the primary objective of a study is model exploration, as in this research, PLS-SEM is considered more appropriate Both theoretically and practically, PLS-SEM functions similarly to multiple regression analysis, with the primary goal of maximizing the explained variance in dependent variables and assessing data quality based on measurement model characteristics Dash G, Paul J (2021). Furthermore, PLS efficiently assesses intricate cause-and-effect relationships. After implementing the PLS algorithm on the research approach, It was assessed for discriminant validity, reliability, and convergence Hair J, et al. (2022). To evaluate dependability and convergence, the following indicators were considered:

• Average Variance Extracted (AVE > 0.50)
• Composite Reliability (CR > 0.70)
• Cronbach's Alpha (Cα > 0.70)

Utilizing the Heterotrait–Monotrait (HTMT) ratio (< 0.85), discriminant validity was examined, which indicates perfect discriminant validity. Additionally, the model's fit quality, path coefficient significance, explanatory power, and predictive relevance were assessed. Evaluating these key indicators helps mitigate the risk of misinterpreting results due to a non-representative sample. Lastly, relationships in the research model were estimated using bootstrapping analysis in Smart PLS (v. 4.0).

## 5. Results
### 5.1. Sociodemographic Characterization of Participants

The final sample consisted of 300 Iraqi consumers aged 18 years and above who had made at least one online purchase from a store utilizing AI-based services within the previous year. Among these respondents, the majority were female (68.25%), with an average age of 27.16 years (ranging from 18 to 60 years). In terms of occupation, 66.4% identified as students, while 29.2% reported being employed.

### 5.2. Statistical Description

Table 1 provides a thorough statistical description of all measured constructs. For constructs related to AI techniques, participants demonstrated the highest levels of agreement with the following dimensions:
• Credibility (M = 4.76), • Entertainment (M = 4.72), • Interaction (M = 4.72). With respect to Consumer Engagement constructs, the highest mean score was observed for Loyalty (M = 4.72), indicating a strong sense of loyalty towards online stores that apply AI techniques.

Similarly, participants showed consistently high agreement levels with items measuring Consumer Purchasing Decision-Making in AI-supported shopping contexts, reflecting a positive influence of AI techniques on their purchase-related decisions.

**Table 1: Statistical Description of construct items**

| Constructs | Items | Mean | Std. d |
|---|---|---|---|
| (AI) techniques | **Entertainment (ENT)** | **4.72** | **0.748** |
| | ENT1 | 4.76 | 0.630 |
| | ENT2 | 4.72 | 0.719 |
| | ENT3 | 4.71 | 0.740 |
| | ENT4 | 4.69 | 0.833 |
| | **Interaction (INT)** | **4.72** | **0.703** |
| | INT1 | 4.74 | 0.700 |
| | INT2 | 4.71 | 0.686 |
| | INT3 | 4.72 | 0.722 |
| | **Credibility (CRE)** | **4.76** | **0.766** |
|  | CRED1 | 4.69 | 0.746 |
|  | CRED2 | 4.67 | 0.779 |
|  | CRED3 | 4.65 | 0.772 |
|  | **Communication (COM)** | **4.64** | **0.769** |
|  | COM1 | 4.69 | 0.713 |
|  | COM2 | 4.61 | 0.824 |
|  | COM3 | 4.64 | 0.771 |
|  | **Customization (CUS)** | **4.69** | **0.757** |
|  | CUS1 | 4.70 | 0.746 |
|  | CUS2 | 4.68 | 0.795 |
|  | CUS3 | 4.67 | 0.783 |
|  | CUS4 | 4.70 | 0.705 |
| Consumer engagement | **Satisfaction (SAT)** | **4.70** | **0.720** |
|  | SAT1 | 4.73 | 0.686 |
|  | SAT2 | 4.72 | 0.706 |
|  | SAT3 | 4.69 | 0.722 |
|  | SAT4 | 4.67 | 0.767 |
|  | **Trust (TR)** | **4.70** | **0.730** |
|  | TR1 | 4.73 | 0.702 |
|  | TR2 | 4.71 | 0.772 |
|  | TR3 | 4.69 | 0.744 |
|  | TR4 | 4.70 | 0.715 |
|  | TR5 | 4.69 | 0.719 |
|  | **Shops loyalty (SL)** | **4.72** | **0.719** |
|  | SL1 | 4.70 | 0.719 |
|  | SL2 | 4.71 | 0.727 |
|  | SL3 | 4.74 | 0.669 |
|  | SL4 | 4.74 | 0.760 |
| Purchasing decision | **Purchase intention (PI)** | **1.95** | **1.438** |
|  | PI1 | 1.98 | 1.385 |
|  | PI2 | 1.92 | 1.478 |
|  | PI3 | 1.95 | 1.453 |

## 5.3 Factor Analysis

The outcomes of the Exploratory Factor Analysis (EFA) and Confirmatory Factor Analysis (CFA) are presented in Table 2. Based on the EFA results, the items were grouped into two distinct factors consistent with the proposed research model.

The cumulative explained variance of these factors reached 63.8%, with no single factor exceeding 50% variance, indicating a balanced distribution among the factors. Furthermore, all items displayed adequate communalities (greater than 0.40), suggesting satisfactory representation within their respective factors. The Confirmatory Factor Analysis results confirmed that all items exhibited significant standardized factor loadings (above 0.50). Accordingly, no item required elimination from the model, thereby supporting the reflective measurement structure proposed in this study

**Table 2: Results of Exploratory and Confirmatory Factor Analyses (EFA, CFA)**

| constructs | items | FACTOR1 | FACTOR2 | Communalities |
|---|---|---|---|---|
| (AI) techniques | **Entertainment (ENT)** | | | |
| | ENT1 | 0.669 | | 0.491 |
| | ENT2 | 0.727 | | 0.680 |
| | ENT3 | 0.722 | | 0.697 |
| | ENT4 | 0.659 | | 0.606 |
| | **Interaction (INT)** | | | |
| | INT1 | 0.701 | | 0.631 |
| | INT2 | 0.719 | | 0.546 |
| | INT3 | 0.644 | | 0.447 |
| | **Credibility (CRE)** | | | |
| | CRED1 | 0.693 | | 0.710 |
| | CRED2 | 0.678 | 0.553 | 0.766 |
| | CRED3 | 0.700 | | 0.731 |
| | **Communication (COM)** | | | |
| | COM1 | 0.721 | | 0.578 |
| | COM2 | 0.614 | | 0.628 |
| | COM3 | 0.653 | 0.510 | 0.503 |
| | **Customization (CUS)** | | | |
| | CUS1 | 0.710 | | 0.710 |
| | CUS2 | 0.623 | | 0.623 |
| | CUS3 | 0.635 | | 0.635 |
| | CUS4 | 0.526 | | 0.526 |
| | **Satisfaction (SAT)** | | | |
| Consumer | SAT1 | 0.753 | | 0.686 |
| engagement | SAT2 | 0.779 | | 0.753 |
| | SAT3 | 0.764 | | 0.701 |
| | SAT4 | 0.570 | 0.559 | 0.637 |
| | **Trust (TR)** | | | |
| | TR1 | 0.705 | 0.502 | 0.749 |
| | TR2 | 0.696 | 0.500 | 0.734 |
| | TR3 | | 0.675 | 0.698 |
| | TR4 | | 0.623 | 0.633 |
| | TR5 | 0.609 | 0.617 | 0.752 |
| | **Shops loyalty (SL)** | | | |
| | SL1 | 0.604 | 0.524 | 0.639 |
| | SL2 | 0.597 | 0.666 | 0.800 |
| | SL3 | 0.679 | 0.557 | 0.771 |
| | SL4 | 0.579 | 0.678 | 0.795 |
| Purchasing decision | **Purchase intention (PI)** | | | |
| | PI1 | | 0.552 | 0.208 |
| | PI2 | | 0.551 | 0.356 |
| | PI3 | | 0.550 | 0.326 |

## 5. 4 Convergence, Validity, and Reliability Tests

The findings of the assessment of the measurement model are summarized in Table 3. The model demonstrates satisfactory convergent validity and reliability, as evidenced by the obtained coefficients exceeding established thresholds: Cronbach's alpha (Cα) > 0.70, Composite Reliability (CR) > 0.70, and Average Variance Extracted (AVE) > 0.50. In terms of discriminant validity, the model meets the criteria since all correlation coefficients among constructs remain below the critical value of 0.85, Benitez J. et al. (2020). Model fit indices further confirm an acceptable fit, with:

• Chi-Square test (p = 0.0762),
• Goodness-of-Fit Index (GFI) of 0.93 (recommended > 0.90),
• Comparative Fit Index (CFI) of 0.91 (recommended > 0.90), and
• Standardized Root Mean Square Residual (SRMR) of 0.057 (recommended < 0.08).

Additionally, the Variance Inflation Factor (VIF) was assessed to detect potential multicollinearity issues. The highest VIF value observed was (3.505) for the item "SL1," which remains below critical thresholds, indicating that multicollinearity does not pose a concern for the model.

**Table 3: Measurement Model Evaluation Results**

|  | Ca | CR | AVE | AI | PDM | SAT | TR | SL |
|---|---|---|---|---|---|---|---|---|
| (AI) techniques | 0.963 | 0.981 | 0.707 |  |  |  |  |  |
| Purchasing decision making (PDM) | 0.758 | 0.871 | 0.758 | 0.730 |  |  |  |  |
| Satisfaction (SAT) | 0.901 | 0.949 | 0.811 | 0.703 | 0.261 |  |  |  |
| Trust (TR) | 0.919 | 0.949 | 0.844 | 0.769 | 0.284 | 0.625 |  |  |
| Shops Loyalty (SL) | 0.927 | 0.927 | 0.859 | 0.740 | 0.245 | 0.598 | 0.667 |  |

The findings regarding the path coefficients' importance and pertinence, evaluated using R² values, are summarized in Table 4. All dependent constructs exhibit a substantial explanatory power,

The R² values indicate the amount of variance in each dependent variable explained by the model. For instance, the R² for Purchasing Decision is 0.533, which means that 53.3% of the variance in purchasing decision-making is explained by the independent variables in the model. Similarly, Satisfaction (0.508), Trust (0.591), and Shops Loyalty (0.548) have moderate to strong explanatory power, indicating that the model accounts for a meaningful proportion of variance in the dependent variables Cohen J (2013).

By computing the Stone-Q2 Geisser's statistic, the structural model's predictive validity was further validated. All of the endogenous constructs' Q2 values were higher than zero, for instance the Q² of (0.422) for Purchasing Decision confirms the model's predictive accuracy in explaining consumer purchasing behavior. Satisfaction (0.398), Trust (0.480), and Shops Loyalty (0.437) Indicating that the model had adequate predictive relevance.

Overall, the high R² and Q² values reflect the robustness and predictive strength of the model, underscoring the important role of AI techniques in enhancing consumer satisfaction, trust, loyalty, and ultimately influencing purchasing decisions within the context of digital marketing.

**Table 4: Significant and predictive relevance**

|  | $R^2$ | $Q^2$ |
|---|---|---|
| Purchasing decision | 0.533 | 0.422 |
| Satisfaction | 0.508 | 0.398 |
| Trust | 0.591 | 0.480 |
| Shops Loyalty | 0.548 | 0.437 |

### 5.5 Estimation of the Research Model

The outcomes of the bootstrapping analysis, used to estimate the hypothesized relationships within the research model, are summarized in Table 5.

The findings indicate that AI techniques exert a positive and highly significant influence on consumer engagement with online shops employing digital marketing strategies. Specifically, the presence of AI techniques was strongly associated with increased levels of consumer loyalty ($\beta$ = 0.982), trust ($\beta$ = 0.972), and satisfaction ($\beta$ = 0.970), thereby supporting Hypotheses H1a, H1b, and H1c.

Additionally, AI techniques were found to have a positive, albeit modest, direct effect on consumer purchasing decision-making ($\beta$ = 0.084), which confirms Hypothesis H2.

The analysis also showed that aspects of consumer engagement had a big impact purchasing decision-making. Among these, trust demonstrated the strongest effect ($\beta$ = 0.094), then loyalty ($\beta$ = 0.087), finally satisfaction ($\beta$ = 0.080). These findings confirm Hypotheses H3a, H3b, and H3c. Additionally, the findings confirmed that consumer engagement had a mediating effect in the relationship between AI techniques and purchasing decisions.

This mediating effect not only assured Hypotheses H4a, H4b, and H4c, but also increased the power of the correlation , refers to influence of AI techniques on purchasing decision-making becomes more strong when consumer engagement is taken into account as a mediator.

Summary these findings highlight that while AI techniques have a direct impact on purchasing decisions, the primary mechanism of influence is through mediating variables satisfaction, trust, and loyalty. This underscores the critical role of consumer engagement factors in translating AI-driven marketing efforts into actual purchase behavior.

**Table 5: Mediating and direct impacts**

| Variables | Path ($\beta$) | t value (Bootstrap) | P values | Confidence interval 2.5% | 97.5% | Hypothesis support |
|---|---|---|---|---|---|---|
| **Direct effects** | | | | | | |
| H1a: AI techniques → satisfaction | 0.970 | 40.557 | 0.000 | -0.828 | 1.111 | yes |
| H1b: AI techniques → trust | 0.972 | 40.754 | 0.000 | -0.833 | 1.112 | yes |
| H1c: AI techniques → Shops Loyalty | 0.982 | 36.834 | 0.000 | -0.086 | 0.264 | yes |
| H2: AI techniques → Purchasing decision | 0.084 | 8.598 | 0.000 | -0.061 | 0.108 | yes |
| H3 a: Satisfaction → Purchasing decision | 0.080 | 8.797 | 0.000 | -0.056 | 0.104 | yes |
| H3 b: Trust → Purchasing decision | 0.094 | 10.960 | 0.000 | -0.072 | 0.116 | yes |
| H3 c: Shop loyalty → Purchasing decision | 0.087 | 10.283 | 0.000 | -0.065 | 0.113 | yes |

| Mediating effects | | | | | | |
|---|---|---|---|---|---|---|
| H4 a: AI techniques → Satisfaction → Purchasing decision | 0.690 | 18.321 | 0.000 | 0.549 | 0.831 | yes |
| H4 b: AI techniques → Trust → Purchasing decision | 0.774 | 21.332 | 0.000 | 0.833 | 1.111 | yes |
| H4 c: AI techniques → Shops loyalty → Purchasing decision | 0.723 | 20.153 | 0.000 | 0.548 | 0.898 | yes |

## 6. Discussion

This study offers insightful information about how AI techniques affect consumer engagement and purchasing decision making in digital marketing environments. The findings support the conclusions of earlier research by showing that AI techniques play an essential role in improving customer satisfaction, enrich trust, and strengthening loyalty, thereby create the overall engagement experience between consumers and shops Lu L et al. (2022).

These results demonstrate the strategic significance of relationship between AI techniques and consumer engagement in digital marketing settings. The consistent positive and significant association observed between AI techniques and purchasing decision-making aligns with Previous research, which have suggested that AI techniques enhances consumers' purchase decisions and influences product selection processes.

The present research emphasizes that the incorporation of AI techniques in digital marketing efforts not only improves consumer engagement but also has a direct effects on consumer' decision-making behavior. Therefore, investing in advanced AI techniques emerges as a successful strategy for merchants looking to influence consumer choices and promote purchase actions.

Additionally, it has been noted that the engagement of consumers with shop that using digital marketing has a positive influence on decision purchases. This beneficial impact on consumers' purchase decisions has been found in a number of consumer engagement characteristics, based on engagements with AI techniques, encompassing the satisfaction, trust, and loyalty Jiang H. et al. (2022).

Moreover, the study found that consumer engagement acts as a role mediating factor in the relevance between AI techniques and purchasing decision. It was discovered that aspects like loyalty, trust, and satisfaction made a substantial contribution to this mediation process. Specifically, trust had the biggest impact on decisions about what to buy, followed by satisfaction and loyalty. According to this, building trust is still crucial in evaluating whether or not customers are ready to make actual purchases, even when AI techniques may draw their attention with tailored services.

The results further emphasize the unique role of every engagement dimension. Trust acts as a psychological bridge between consumers and shops utilizing digital marketing, boosting trust in the veracity of the digital platforms as well as the information supplied by AI techniques. Loyalty, on the other hand, shows a consumer's sustained dedication, which positively affects the intention to repeat purchases. Satisfaction demonstrated a relatively lower direct effect, which may be explained by the fact that it is cumulative nature requiring consistent positive experiences over time, as opposed to loyalty and trust, which appear frequently as instant reactions to satisfying experience Balakrishnan J. (2018).

An important contribution of this study is the demonstrating how consumer engagement has a positive mediating effect, which amplifies the impact of AI techniques on consumer decision-making. When AI-driven services match with consumer expectations and preferences, they facilitate a more satisfying and personalized interaction, increasing consumers' willingness to engage with and purchase from shops that use digital marketing. This underscores how crucial it is to use AI as a tool to improve relationship quality and customer-centric value creation, not only as a tool technical one.

However, there are serious worries about possible future hazards when AI techniques is used in business-to-consumer engagement. There are concerns that these techniques may outsmart humans and have an impact on decision-making because of their extraordinary speed and capacity to retain large volumes of behavioral data Pelau et al. (2021). Overuse of AI techniques can increase the possibility of consumer deception, can cause Accreditation on these techniques damage social connections and decrease cognitive capacities, change personality processes and thought.

**6.1 Theoretical implications**

This study adds significantly benefactions to the corpus of research with regard to the role of AI techniques- enabled in the context of digital marketing.

First, the results reinforce the idea that AI techniques in digital platforms impact consumer impressions and engagements. This emphasizes the requirement for Researchers to more research the importance of AI techniques in enriching consumer satisfaction, trust, and loyalty within AI-powered digital marketing frameworks.

Second, the investigation fosters the knowing of consumer engagement construct as a mediator in the correlation between AI techniques and purchasing decision-making.

It shows that consumer loyalty, satisfaction, and trust are crucial in strengthening the influence of AI techniques on purchasing decision. This result confirms the requirement for deeper theoretical search of affection and correlational factors as important leaders of purchasing decision in AI-Supported engagements. The investigation indicates consumer engagement is more than just a result of satisfactory experiences but also works as a style during which AI techniques impact purchase decision-making also, merits scholars' interest.

Third, the Verify of the study framework and the amplification of current theories supply a further precise viewpoint on the styles by which artificial intelligence techniques impact consumer behavior, consequently creating new opportunities for further research.

Fourth, the theoretical framework this study offers makes it easier to apply the Basic constructs inside the framework of AI techniques and consumer engagement, providing a better understanding of how these constructs function in AI-driven engagement. Because they can be used as a key to further support and develop these constructions, these results are worthy of academics' attention and constitute a useful addition to continuing theoretical research.

Finally, the findings suggest that careful ethical consideration is necessary, as the impact of AI techniques on consumer engagement and purchase decision raises concerns regarding the possible for Artificial intelligence techniques to tampering consumer behavior. Academics need to look into the subtle ways that AI techniques could influence decisions or promote purchases without the customers completely understanding the persuasive strategies being employed. Academia have other ethical subject that require Research, especially with regard to transparency. Furthermore, certain consumer segments, especially those who are unfamiliar with artificial intelligence or digital technology, might be more vulnerable to the impact of AI techniques, and this issue also needs additional scholarly investigation.

**6.2 Implications for practice**

Implications for Managers:

Practically speaking, managers can gain a great deal of insight from this study.

First, according to the research findings, correctly implemented AI techniques have the ability to improve consumer experiences, streamline processes, and open up new business prospects. As a result, businesses should welcome and strategically benefit from the AI revolution, since ignoring it or putting it into practice too quickly could have serious consequences Xiao L, Kumar V (2021). Second, since this study shows that the AI techniques linked to digital marketing greatly boost consumer loyalty, satisfaction, trust and to shops that use it , in addition impact purchasing decisions, it is critical for businesses to incorporate the all AI techniques into the planning stages of their digital marketing programs .

Thirdly, the study emphasizes the advantages of AI techniques. In improving consumer experience and accomplish important objectives like retention and conversion, businesses need invest in cutting-edge designs and sophisticated communication tools that boost engagement and guarantee smooth interactions.

Fourthly, managers should take into account the substantial impact of AI techniques on consumer responses and behaviors Dong et al. (2020). When developing digital marketing tools, as research has demonstrated that these techniques are essential in influencing consumer behavior.

Fifthly, enhancing consumer engagements with Ai techniques necessitates striking an equilibrium between tech, customization, and consumer experiment. In order to achieve this, digital shops able increase the impact of AI techniques by using more natural, approachable language tailored to the target consumers ; responding to consumer issues with greater empathy; using humanized voices or avatars that represent the shop without resorting to exaggerations that might inspire mistrust; personalizing answers to increase the consumer's sense of value, enhancing the consumer experiment; ensuring transparency and lucidity, alerting consumers from the outset that they are engagement with AI techniques, preventing embarrassment, and offering channel varied employment; merge the AI techniques into Programs, applications ,networks, and websites and guaranteeing a Perfect experiment through every channel.

Lastly, the knowledge gained from this study on AI techniques may be extended to a number of fields outside of digital marketing, providing helpful advice on how diverse sectors might use this technology to improve consumer engagement and affect adoption or buy intentions. For example, in the fields of public services, education, and healthcare, as well as banking and finance. This improves the public's opinion of the government and increases digital inclusion for people who are not very tech-savvy.

Implications for Customers:

The findings show that AI techniques have a favorable impact on consumers' intention to buy and their level of engagement with online shops. From these findings, the following practical implications for consumers can be deduced: I a stronger emotional connection to the shopping experience; (ii) the availability of AI techniques that can lessen consumers' anxiety or uncertainty while browsing; (iii) AI techniques that employ customization give the impression is distinctive and tailored to the needs of the client; (iv) consumers will feel as though their questions or issues are answered more quickly and clearly when engaging with AI techniques; (v) AI techniques can make customers think that the digital shop is more dependable and readily available ; (vi) using AI techniques can change the shopping experience into something more satisfactory.

With all of these practical implications, it is important to critically consider the ethical issues raised by the application of AI techniques. These issues include the erosion of privacy, consumer manipulation, the reduction of human agency, and the reinforcement of biases.

### 6.3. Limitations

While this study provides valuable insights into the role of AI techniques in shaping consumer engagement and purchasing decision-making, certain limitations should be acknowledged. First, the findings are derived from a non-probabilistic convenience sample composed primarily of young Iraqi consumers (average age 27.16 years), the majority of whom are students without full time professional occupations. As such, the results may not be generalizable to the broader population of Iraqi consumers with more diverse demographic and professional backgrounds.

Second, the sampling approach may have introduced selection bias, as participation was voluntary and likely influenced by the author's social network reach. This may have resulted in the overrepresentation of certain consumer segments while underrepresenting others, such as older individuals or those with less exposure to digital platforms.

Third, this research focused on only five functional characteristics of AI techniques entertainment, interaction, credibility, communication, and customization. Other potentially relevant attributes, such as competence, sensitivity, confidence, emotional expression, and commitment, were not included and could affect the conceptual model's explanatory power if integrated into future studies.

Additionally, consumer engagement was assessed exclusively through the constructs of satisfaction, trust, and loyalty. Although these are well-established dimensions, other aspects such as emotional attachment, perceived value, or relationship quality might provide a more comprehensive understanding of engagement in AI-driven contexts.

Finally, the measurement of purchasing decision-making was limited to three items. Although factor analysis confirmed the adequacy of these indicators in capturing the construct, the inclusion of additional items such as preference for AI-equipped shops over non-AI alternatives, willingness to increase purchase frequency, and intention to recommend AI-driven shops could enhance the robustness and validity of this dimension.

These limitations emphasize the need for more research using probabilistic sampling techniques, larger consumer segments, and wider measurement scales to validate and extend the study's conclusions and advise caution when interpreting and extrapolating the results.

### 6.4 Future research directions

The link between consumer engagement and purchase decision-making should be moderated in future research by using sociodemographic factors (e.g., age, gender, education, and income) and obtaining a representative probability sample.

Shop loyalty and preference may also be employed as customer engagement factors in future research. Additionally, future research would involve identifying a particular store that employs

AI techniques and investigating the features of those techniques as well as the experience, interaction, and purchasing decisions of that shop's clientele.

Future study might benefit from examining how past experiences with AI techniques function as a mediator in the correlation between consumer engagement and purchase decisions. Studying AI techniques in various cultural contexts or with various technical interfaces would also be crucial. As a result, it would be crucial, for example, to assess which aspects of an AI techniques (language, appearance, and voice) are optimal approvable in difference cultures and whether there is a diversity in the style difference cultures refers to intentions or emotions to AI techniques.

Furthermore, given the multiplicity of technological interfaces, it would be crucial to investigate, for instance, how different written, visual, and auditory content interfaces perceive techniques; whether multimedia engagement the mixture of voice, visual, text raise or reduce the influence of techniques; and if sophisticated interfaces have a greater impact on customers' feeling or trust .

Given the variations in Products types, it should be crucial for following research to examine if the effects of AI techniques differ based on the kind of service (entertainment, banking, health care ) or goods (long-lasting versus short-lived  consumer goods)  and determination which stages of marketing are most work best for various product categories.

## 7. Conclusion

The growth of shops employing AI techniques has transformed consumer behavior and decision-making within digital marketing environments. This study investigates how AI techniques affect consumers' engagement with shops that employ digital marketing and how they purchasing decision-making. The findings show that greater usage of AI techniques greatly boosts consumer satisfaction, trust, and loyalty to shops that employ them. Additionally, it influences consumers' purchase decisions, albeit in a residual way. The results show that consumer engagement plays a crucial mediating role, intensifying the relationship between AI techniques and Consumer purchasing decision. The relationship between AI techniques and consumer purchasing decision is increases when it is mediated by consumer engagement with shops that utilize digital marketing. So, it is essential to Strengthening satisfaction, trust and loyalty to the shop, for AI techniques to more important impact consumer purchasing decision.

In summary, the incorporation of AI techniques into digital marketing presents considerable potential to changing consumer experiences and influence decision-making processes. However, realizing this potential requires a strategic focus on building consumer relationships characterized by trust, satisfaction, and loyalty key factors that determine the success of AI-driven marketing initiatives.

**Appendix A.1 – Questionnaire**

The full questionnaire used for data collection in this study is available at:
https://docs.google.com/document/d/1-rfIjUaW6qB1SCtybyANmevAu_xpR9w9/edit